\documentclass[journal]{IEEEtran}

\usepackage{url}
\usepackage{graphicx}
\usepackage{booktabs}
\usepackage{multirow}
\usepackage{amsfonts}
\usepackage{wasysym}
\usepackage{xcolor}
\usepackage{titlesec}

\titlespacing*{\subsection}{0pt}{0.5\baselineskip}{0.5\baselineskip}
\titlespacing*{\section}{0pt}{1.0\baselineskip}{0.7\baselineskip}

\begin{document}

\title{Multi-target Extractor and Detector for Unknown-number Speaker Diarization}

\author{
Chin-Yi Cheng, Hung-Shin Lee, Yu~Tsao,~\IEEEmembership{Senior~Member,~IEEE}, and Hsin-Min Wang,~\IEEEmembership{Senior~Member,~IEEE}}

\maketitle

\begin{abstract}
Strong representations of target speakers can help extract important information about speakers and detect corresponding temporal regions in multi-speaker conversations. In this study, we propose a neural architecture that simultaneously extracts speaker representations consistent with the speaker diarization objective and detects the presence of each speaker on a frame-by-frame basis regardless of the number of speakers in a conversation. A speaker representation (called z-vector) extractor and a time-speaker contextualizer, implemented by a residual network and processing data in both temporal and speaker dimensions, are integrated into a unified framework. Tests on the CALLHOME corpus show that our model outperforms most of the methods proposed so far. Evaluations in a more challenging case with simultaneous speakers ranging from 2 to 7 show that our model achieves 6.4\% to 30.9\% relative diarization error rate reductions over several typical baselines.
\end{abstract}


\IEEEpeerreviewmaketitle

\section{Introduction}

\IEEEPARstart{S}{peaker} diarization (SD) is the process of determining when individual speakers are active in a recording. The aim is to generate a diary of the presence of each speaker at each point in time. This technique has been widely used in speech processing in various scenarios, such as conversations, broadcast news, debates, and cocktail parties \cite{Anguera2010}. However, the robustness of SD remains weak due to the challenges posed by the changes in recording channel, environment, reverberation, ambient noise, and number of speakers \cite{Ryant2021}.

Researchers have tackled SD tasks using probabilistic models \cite{Diez2018a} or neural networks (NNs) \cite{Fujita2019}. Many methods involve two steps, segmentation and clustering. In the segmentation step, a session is divided into a series of short segments, typically using a sliding window of 1.5 seconds with 50\% overlap. Then, a speaker model is used to extract the speaker representation (e.g., the x-vector \cite{Snyder2018,Daniel2017,Diez2019,Sell2018}, i-vector \cite{Dehak2011, Sell2014}, or d-vector \cite{Wan2018,Zhang2018}) for each segment. During clustering, segments with homogeneous characteristics form a group. Many clustering techniques have been used based on various similarity measures, such as probabilistic linear discriminant analysis (PLDA) and cosine similarity \cite{Ioffe2006,Simon2007,Kenny2013,Sell2014}. For example, agglomerative hierarchical clustering (AHC) and spectral clustering (SC) were used in \cite{Daniel2017,Sell2014} and \cite{Ning2006,Park2020}, respectively. The unbounded interleaved-state recurrent neural network (UIS-RNN) derived from the Gaussian mixture model (GMM) \cite{Bozonnet2010,Shum2013} and hidden Markov model (HMM) \cite{Diez2019} was used in \cite{Zhang2018}. Moreover, some post-processing methods, such as variational Bayes (VB) \cite{Sell2015} and the LSTM-based method \cite{Sahidullah2019}, have been used to refine the initial SD results.

Several recent studies \cite{Fujita2019b,Neumann2018,Fujita2019d,Bullock2020} have focused on end-to-end SD. Fujita \textit{et al.} \cite{Fujita2019b} reformulated the SD task as a multi-label classification problem and used the permutation-invariant training (PIT) \cite{Yu2017} technique. In \cite{Zeghidour2021}, reliable speaker representations were derived using a selector to assist the voice activity detector in diarizing a session. Self-attention \cite{Fujita2019c,Horiguchi2020} and frame selection \cite{Horiguchi2021} have also been used in end-to-end SD.

Traditional segmentation-clustering methods cannot handle overlapping speech in a session. Target speaker voice activity detection (TS-VAD) \cite{Medennikov2020a} can handle overlapping speech well. It relies on the x-vector/SC method \cite{Snyder2018,Sell2018} to provide initial speech regions for each ``target'' (active) speaker, and then extracts his/her i-vector from the corresponding frames. Finally, it uses the i-vectors of all speakers and MFCCs to generate the SD result. Unfortunately, it can only be applied to sessions with a fixed number of speakers due to a tensor concatenation of speaker representations. Inspired by the dual-path recurrent neural network (DPRNN) \cite{Luo2020,Chenda2021}, we propose a unified structure called Multi-target Extractor and Detector (MTEAD), which can handle sessions with various numbers of speakers using a single model. Furthermore, since the quality of speaker representations has an impact on SD performance, we extend TS-VAD by using a neural extractor that can extract speaker representations suitable for SD. Note that in our framework, the number of speakers in a test session is determined by the initial x-vector/AHC SD process.

Our main contributions are threefold. First, we extend the practical scope of TS-VAD while inheriting its excellent SD performance. Second, we design an extractor that can be jointly trained with the SD model to extract speaker representations. Using this extractor, we achieve better SD performance and avoid the pretraining of the i-vector extractor. Third, unlike some previous studies \cite{Daniel2017,Zhang2018} that only dealt with non-overlapping speech, our model can handle overlapping speech.


\begin{figure}[t]
\centering
\begin{center}
\includegraphics[width=0.33\textwidth]{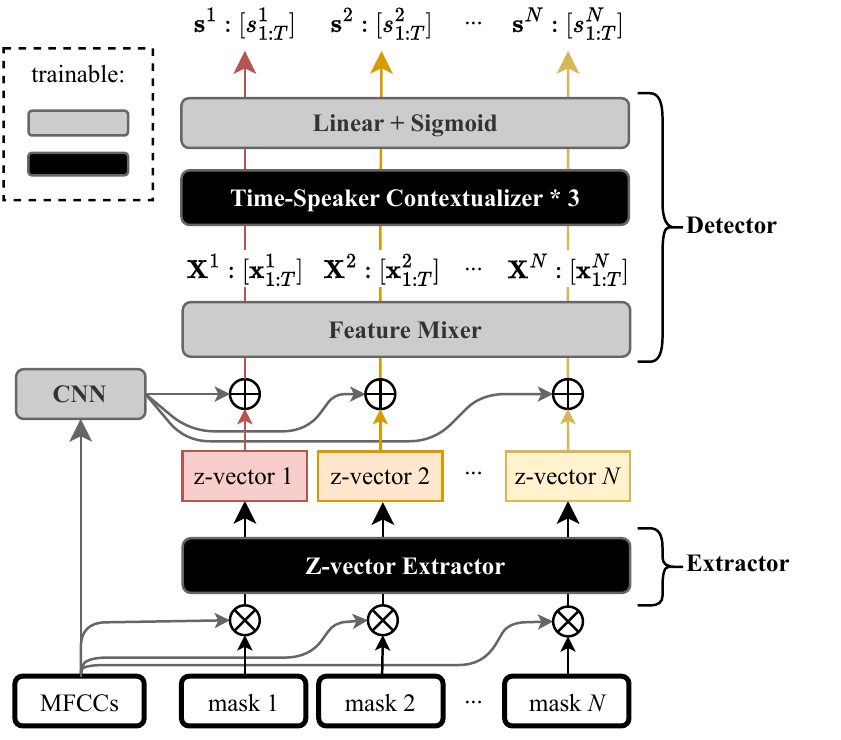}
\end{center}
\vspace{-10pt}
\caption{Structure of MTEAD. $\oplus$ denotes the concatenation operator. $\mathbf{X}^i$ and $\mathbf{s}^i$ are the $\rm S_{d}$ frames and diarization scoring vectors for speaker $i$, respectively. $\otimes$ is an element-wise product operator of MFCCs, and the binary masks (speaker occurrence labels) are obtained in the first step (x-vector/AHC diarization).}
\label{fig:MTEAD}
\vspace{-15pt}
\end{figure}

\section{Multi-target Extractor and Detector}


\subsection{Framework}\label{sec:framework}
Inspired by TS-VAD \cite{Medennikov2020a}, we propose a two-step SD method called MTEAD, which can be adapted to different numbers of speakers. MTEAD not only addresses the main weakness of TS-VAD (i.e., handling only sessions with a fixed number of speakers, e.g., four in \cite{Medennikov2020a}), but also uses improved speaker representations. As shown in Fig. \ref{fig:MTEAD}, the first step of MTEAD (i.e., Extractor) relies on a traditional x-vector/AHC SD method to initially detect the speech regions for each speaker (i.e., speaker occurrence labels). Then, the frames corresponding to each speaker are used to extract his/her speaker representation. Here, the representation may be a traditional i-vector or x-vector, or our specially designed z-vector for SD. The second step of MTEAD (i.e., Detector) requires two inputs: the frame-level MFCCs and the representation of each speaker. The frame-level MFCCs are first passed through a four-layer convolutional neural network (CNN), and then concatenated with each speaker representation. The output of Detector is the final diarization result for each speaker.

In TS-VAD, the second step is implemented using BiLSTM. The first two layers of the BiLSTM take the frame-level MFCCs and four i-vectors as input, and output four speaker detection ($\rm S_{d}$) vector sequences. Then, the four $\rm S_{d}$ vector sequences are concatenated along the feature dimension and passed through the third layer of the BiLSTM to generate the final diarization result for each speaker. It is this concatenation that causes TS-VAD to only handle four-speaker recordings. Furthermore, it uses the i-vector as the speaker representation, which is defeated by the x- and z-vectors in our experiments.


\subsection{Detector: Feature Mixer / Time-Speaker Contextualizer}\label{sec:detector}
\begin{figure}[t]
\centering
\includegraphics[width=0.46\textwidth]{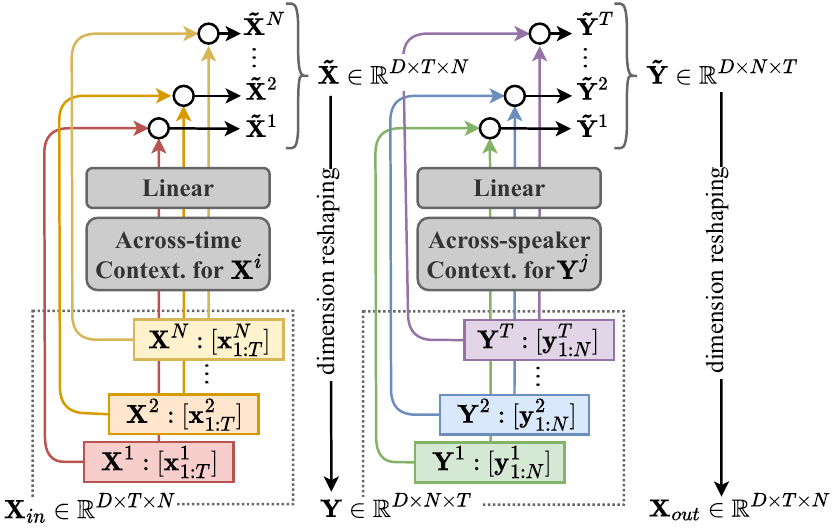}
\vspace{-5pt}
\caption{The Time-Speaker Contextualizer. The operator $\ocircle$ denotes the residual addition of two tensors. $\mathbf{X}^i$ and $\mathbf{Y}^j$ are passed through the Across-time and Across-speaker Contextualizer, respectively.}
\vspace{-15pt}
\label{fig:frame_speaker}
\end{figure}

The Detector consists of a Feature Mixer and three consecutive Time-Speaker Contextualizers implemented using BiLSTM. The MFCCs are passed through CNNs and concatenated with each speaker representation. These concatenated features are then processed by Feature Mixer to generate the corresponding $\rm S_{d}$ frames $\mathbf{X}^i$ for the $i$-th speaker, as shown in Fig.~\ref{fig:MTEAD}. As illustrated in Fig.~\ref{fig:frame_speaker}, the stack of the $\rm S_{d}$ frames of all speakers, $\mathbf{X}_{in} \in \mathbb{R}^{D\times T\times N}$, is the input of the Time-Speaker Contextualizer, where $D$, $T$, and $N$ denote the dimension of the $\rm S_{d}$ vector, number of frames, and number of speakers, respectively. Inspired by DPRNN, the Time-Speaker Contextualizer is designed to handle different numbers of speakers in one session. It contains two stages. In the first stage, the $\rm S_{d}$ vector sequence of each speaker is passed through the Across-time Contextualizer separately to generate the temporal contextual information using
\begin{equation}
\tilde{\mathbf{X}}^i = Linear(Contextualizer_T(\mathbf{X}^i)) + \mathbf{X}^i.
\end{equation}
The output of the first stage, $\tilde{\mathbf{X}}$, is the stack of $\tilde{\mathbf{X}}^i$, $i=1,...,N$. The input of the second stage, $\mathbf{Y}$, is generated by reshaping $\tilde{\mathbf{X}}$. Slicing the input by timeframes yields $\mathbf{Y}^j \in \mathbb{R}^{D\times N}$, $j=1,...,T$, which is treated as the activity of individual speakers in a single frame $j$. Then, $\mathbf{Y}^j$ is passed through the Across-speaker Contextualizer to generate the speaker contextual information using
\begin{equation}\label{EQ:2}
\tilde{\mathbf{Y}}^j = Linear(Contextualizer_S(\mathbf{Y}^j)) + \mathbf{Y}^j.
\end{equation}
$\tilde{\mathbf{Y}}^j$, $j=1,...,T$, is stacked and reshaped to $\mathbf{X}_{out} \in \mathbb{R}^{D\times T\times N}$. $\mathbf{X}_{out}$ is used as the input, $\mathbf{X}_{in}$, of the subsequent Time-Speaker Contextualizer. Finally, for each speaker, the corresponding $\rm S_{d}$ vector sequence from the output of the last Time-Speaker Contextualizer is passed through a linear-sigmoid layer to generate the final diarization result.

In Eq. (\ref{EQ:2}), the number of speakers $N$ is the length of the input sequence; therefore, it is variable. The Across-speaker Contextualizer allows the sharing of speakers' information within a single frame, which can help Detector to discern differences among speaker representations. On the other hand, by scanning along the frame, the Across-time Contextualizer can help Detector to determine whether each speaker is active in each frame. Therefore, MTEAD achieves information sharing among all speakers and frames through the Across-speaker and Across-time Contextualizers, which not only preserves the advantages of TS-VAD, but also does not have the limitation of handling only conversations of a fixed number of speakers.

\subsection{Extractor: z-vector (diari``z"ation vector)}\label{sec:z-vector}

We argue that the quality of speaker representations is critical to SD performance. Therefore, we design an extractor specifically for extracting speaker representations suitable for SD. As shown in Fig. \ref{fig:MTEAD}, using the speaker speech regions provided by x-vector/AHC diarization and the MFCCs of the session, the Z-vector Extractor generates the corresponding speaker representations\footnote{While we were preparing this paper, there was a similar work extending TS-VAD by jointly training a speaker embedding network \cite{Wang2022}. However, it does not handle a variable number of speakers.}. The Z-vector Extractor comprises ResNet and Attentive Statistics Pooling (ASP) \cite{Okabe2018}. The ResNet here is similar to ResNet50, but the last layer is replaced by the ASP layer. The input to ResNet is a sequence of 500 frames of MFCCs from each speaker, and the output shape is set to be the same dimension as the x-vector\footnote{The number of parameters for the x- and z-vector extractors is about 8M and 30M, respectively.}. Therefore, the z-vectors can be used as input to Detector instead of x- or i-vectors. In MTEAD, speaker representation extraction and speaker detection are combined into a unified process by integrating Extractor with Detector. Also, since Extractor and Detector are trained jointly, it is expected that the z-vector is more suitable for SD than the x- and i-vectors.

\subsection{Joint training and loss}\label{sec:loss}
When using z-vectors, Extractor and Detector are jointly trained from scratch using a binary cross-entropy loss between speaker labels and predicted SD results. The per-speaker losses are summed directly and back-propagated. When using x-vectors and i-vectors, Extractor is replaced by other extractors pretrained by the Kaldi recipe, and only Detector is trained using the same binary cross-entropy loss. 

\begin{table}
\centering
\caption{DER (\%) of MTEAD with x-vector on the SWB+SRE dataset.}
\vspace{-5pt}
\label{tab:exp}
\begin{tabular*}{\linewidth}{@{\extracolsep{\fill}} cccc}
\toprule
\multicolumn{1}{c}{\textbf{Ratio} ({$|G|/(|G|+|L|)$}}) & \textbf{Threshold} & \textbf{Oracle} & \textbf{Ideal} \\
\midrule
\midrule
0\% & 29.92 & 26.37 & \textbf{16.65} \\
25\% & 29.82 & 25.91 & \textbf{12.03} \\
50\% & 30.05 & 26.26 & \textbf{11.14} \\
75\% & 29.80 & 26.22 & \textbf{10.36} \\
100\% & 30.43 & 27.74 & \textbf{10.22} \\
\bottomrule
\end{tabular*}
\vspace{-15pt}
\end{table}
\section{Experiments and Results}
\vspace{-5pt}
Two corpora were used in our experiments: one was simulated from the Switchboard and NIST SRE datasets, and the other was CALLHOME. The results were evaluated by diarization error rate (DER) and Jaccard error rate (JER) \cite{Hamers1989}\cite{Real1996}\cite{Ryant2019}, with a standard 250 ms collar. Overlapping speech frames were counted. Both Feature Mixer and Across-time Contextualizer were implemented using BiLSTM. Across-speaker Contextualizer can be implemented using Transformer \cite{Vaswani2017} or BiLSTM. As the number of speakers is limited in both datasets (i.e., less than 10), a lightweight BiLSTM is sufficient to gather all information about the speakers. For model training, we used the Adam optimizer and Noam scheduler with a learning rate of $0.1$ and a batch size of $8$. The dimensions of i-vector, x-vector, and z-vector were all set to 40\footnote{See https://github.com/chinyi0523/MTEAD for details.}.

\subsection{SWB+SRE Simulated Corpus}
The total number of speakers in the 683 hours of data from the SRE and Switchboard datasets was 6,392. We simulated the training and evaluation data by Algorithm 1 in \cite{Fujita2019c}. The speakers in the two sets did not overlap. To achieve 20\% overlap, for the cases of two, three, and four speakers, the parameter $\beta$ in the algorithm was set to 3, 6, and 9 seconds, resulting in 137, 226, and 320 hours of data, respectively.

During training, the ground truth in Rich Transcription Time Marked (RTTM) format was used as the speaker occurrence labels for each session. During inference, we used x-vector/AHC to produce the RTTM file for each test session. The \textit{Threshold} in x-vector/AHC was set based on the performance on the training set. Generation of speaker representations followed the methods described in Secs. \ref{sec:framework} and \ref{sec:loss}. 

\textbf{Results and discussion}. First, we investigated the impact of the quality of speaker representations on SD performance. Extractor in MTEAD (cf. Fig. \ref{fig:MTEAD}) was replaced with a pretrained x-vector extractor. During MTEAD training, each target speaker's x-vector was extracted from all his/her original utterances in Switchboard and NIST-SRE (denoted as G (global)) or from the corresponding speech frames in a session based on the ground truth RTTM (denoted as L (local)). The ratio $|G|/(|G|+|L|)$ indicates the degree to which global speaker representations were used for training. The \textbf{Threshold} and \textbf{Oracle} tests denote that the number of clusters in AHC is automatically determined and set to the true number of speakers, respectively. \textbf{Ideal} denotes the (cheating) case where the target speaker's x-vector was extracted from the test session based on the ground truth RTTM. As shown in Table \ref{tab:exp}, the ratio has little impact on SD performance, showing that x-vectors extracted from a limited amount of speech in a training session are as effective as x-vectors extracted from a large amount of speech in the entire training dataset. Regardless of the ratio, all \textbf{Ideal} test conditions outperformed their \textbf{Threshold} and \textbf{Oracle} counterparts. These results demonstrate the importance of accurate speaker representations, showing that extracting better speaker representations is key to producing excellent SD results. 

Next, we compared the performance of different speaker representation models, including z-vector, i-vector, and x-vector. As shown in Table \ref{tab:SWBSRE_Result}, while the MTEAD model with i-vectors or x-vectors showed better performance than the baseline x-vector/AHC method, the MTEAD model with z-vectors achieved the best performance. The results confirm that the z-vector jointly trained with the MTEAD model is more effective than the x-vector and i-vector obtained from pretrained models. Speaker representations targeting SD did yield better SD performance. 

\begin{table}[t]
\centering
\caption{Results (\%) on the SWB+SRE dataset. All MTEAD models are based on the initial diarization of x-vector/AHC.}
\vspace{-5pt}
\label{tab:SWBSRE_Result}
\begin{tabular*}{\linewidth}{@{\extracolsep{\fill}} lcccc}
\toprule
\multicolumn{1}{c}{\multirow{2}{*}{\textbf{Method}}} & \multicolumn{2}{c}{\textbf{Threshold}} & \multicolumn{2}{c}{\textbf{Oracle}}\\
\cmidrule{2-3}
\cmidrule{4-5}
 & DER & JER & DER & JER \\
\midrule
\midrule
x-vector/AHC & 38.49 & 53.38 & 40.38 & 52.58 \\
\midrule
MTEAD (i-vector) & 23.72 & 36.21 & 19.50 & 29.24 \\
MTEAD (x-vector) & 25.16 & 38.91 & 18.61 & 28.88 \\
MTEAD (z-vector) & \textbf{23.06} & \textbf{32.67} & \textbf{13.46} & \textbf{18.95} \\
\bottomrule
\end{tabular*}
\vspace{-15pt}
\end{table}

\subsection{CALLHOME (LDC2001S97)}

CALLHOME is a telephony dataset containing conversations in multiple languages. The dataset consists of a total of 500 conversations recorded at a sampling rate of 8 kHz. The number of speakers in each conversation varies from 2 to 7. Since the CALLHOME dataset is too small to train our model, we used the SWB+SRE dataset for pretraining.

During training, we pretrained the MTEAD models on SWB+SRE. We divided the CALLHOME set equally into two parts as Kaldi instructed. CALLHOME-1 was used for fine-tuning the pretrained models, while CALLHOME-2 was used for evaluation. We determined the \textit{Threshold} in x-vector/AHC based on the performance on CALLHOME-1. During inference, we used x-vector/AHC to produce RTTM files on CALLHOME-2. We used the same methods as the experiments on SWB+SRE to generate speaker representations.

As the number of speakers in a CALLHOME session varies from 2 to 7, it is necessary to pretrain and fine-tune a TS-VAD model for each speaker number separately. To this end, the SWB+SRE and CALLHOME-1 datasets were split into 2-, 3-, and 4-speaker subsets for training the corresponding TS-VAD models. For each TS-VAD model, we also trained the corresponding MTEAD* model using the same training data and procedure for fair comparison. MTEAD was trained on all training data containing different numbers of speakers. 

\begin{table}[t]
\centering
\caption{Results (\%) on CALLHOME based on i-vectors}
\vspace{-5pt}
\label{tab:CallHome_Result_TS}
\begin{tabular*}{\linewidth}{@{\extracolsep{\fill}} lcccccc}
\toprule
\multicolumn{1}{@{\hspace{0pt}}c}{\multirow{2}{*}{\textbf{Method}}} & \multicolumn{2}{@{\hspace{0pt}}c}{\textbf{Oracle \#2}} & \multicolumn{2}{@{\hspace{0pt}}c}{\textbf{Oracle \#3}} & \multicolumn{2}{@{\hspace{0pt}}c}{\textbf{Oracle \#4}} \\ 
\cmidrule{2-3}
\cmidrule{4-5}
\cmidrule{6-7}
& DER & JER & DER & JER & DER & JER\\
\midrule
\midrule
\multirow{2}{*}{\shortstack[1]{SA-EEND\\-EDA \cite{Horiguchi2020}}} & \multirow{2}{*}{8.35} & \multirow{2}{*}{N/A} & \multirow{2}{*}{13.20} & \multirow{2}{*}{N/A} & \multirow{2}{*}{21.71} & \multirow{2}{*}{N/A} \\
&&&&&& \\
\midrule
x-vector/AHC & 9.17 & 24.94 & 15.24 & 37.04 & 20.28 & 45.35 \\
\midrule
TS-VAD & 9.51 & 20.60 & 14.71 & 33.62 & 20.18 & 44.77\\
MTEAD* & 8.72 & 17.90 & 14.50 & 33.60 & 18.15 & 43.24\\
MTEAD & \textbf{7.82} & \textbf{17.87} & \textbf{13.10} & \textbf{32.43} & \textbf{18.12} & \textbf{39.02}\\
\bottomrule
\end{tabular*}
\vspace{-5pt}
\end{table}

\begin{table}[t]
\centering
\caption{Results (\%) on CALLHOME. All MTEAD models were based on the initial diarization of x-vector/AHC.} 
\vspace{-5pt}
\label{tab:Callhome_Result_all}
\begin{tabular*}{\linewidth}{@{\extracolsep{\fill}} lcccc}
\toprule
\multicolumn{1}{c}{\multirow{2}{*}{\textbf{Method}}} & \multicolumn{2}{c}{\textbf{Threshold (Estimated)}} & \multicolumn{2}{c}{\textbf{Oracle}}\\
\cmidrule{2-3}
\cmidrule{4-5}
 & DER & JER & DER & JER \\
\midrule
\midrule
x-vector/AHC \cite{Huang2020} & 20.71 & N/A & 20.14 & N/A \\
x-vector/AHC+VB \cite{Huang2020} & 19.51 & N/A & 18.61 & N/A \\
SA-EEND-EDA \cite{Horiguchi2020} & 15.29& N/A & 15.43& N/A\\
\midrule
MTEAD (i-vector) & 14.52 & 30.09 & 14.10 & 27.92 \\
MTEAD (x-vector) & 14.55 & 30.01 & 13.15 & 26.80 \\
MTEAD (z-vector) & \textbf{14.31} & \textbf{29.21} & \textbf{12.66} & \textbf{24.56} \\
\bottomrule
\end{tabular*}
\vspace{-15pt}
\end{table}

\begin{table}[t]
\caption{DER (\%) on the 2-speaker CALLHOME task.}
\vspace{-10pt}
\label{tab:CallHome_2speaker}
\center
\begin{tabular*}{\linewidth}{@{\extracolsep{\fill}} lcc}
\toprule
\multicolumn{1}{c}{\textbf{Method}} & \textbf{DER} & \textbf{rel. \%}\\
\midrule
\midrule
x-vector/AHC \cite{Horiguchi2020} & 8.93 & - \\
\midrule
BLSTM-EEND \cite{Fujita2019d} & 23.07 & -158.3 \\
SA-EEND \cite{Fujita2019c} & 10.99 & -23.1 \\
SA-EEND-EDA \cite{Horiguchi2020} & 8.35 & 6.5 \\
SA-EEND-EDA + Frame Selection \cite{Horiguchi2021} & 7.84 & 12.2 \\
DIVE \cite{Zeghidour2021} & 6.70 & 24.9\\
\midrule
MTEAD & 7.82 & 12.4 \\
\bottomrule
\end{tabular*}
\vspace{-15pt}
\end{table}

\textbf{Results and discussion}. First, we compared MTEAD with TS-VAD. TS-VAD, MTEAD*, and MTEAD were all based on the initial diarization of x-vector/AHC, and were all implemented with i-vectors. For each specific number of speakers, MTEAD* and TS-VAD were trained on the same training data. The experiments were conducted under the 2-, 3-, and 4-speaker conditions, assuming the number of speakers was known. The evaluation results on CALLHOME-2 are shown in Table \ref{tab:CallHome_Result_TS}. It is clear that MTEAD* consistently outperformed the corresponding TS-VAD and x-vector/AHC baselines under each identical condition. Moreover, MTEAD outperformed MTEAD* because it was trained with all training data containing different numbers of speakers, while each MTEAD* model was trained with only a subset of training data with a specific number of speakers. Results demonstrate the advantage of the MTEAD Detector: MTEAD significantly outperformed TS-VAD because its detector could be trained using all training data with different numbers of speakers. After overcoming the weakness of TS-VAD, MTEAD with i-vector has already surpassed SA-EEND-EDA \cite{Horiguchi2020}.

Next, we compared the performance of different speaker representation models, including z-vector, i-vector, and x-vector. As can be seen from Table \ref{tab:Callhome_Result_all}, all three MTEAD models outperformed not only the x-vector/AHC and x-vector/AHC+VB baselines, but also the strong SA-EEND-EDA model \cite{Horiguchi2020}. Furthermore, z-vector-based MTEAD outperformed i-vector-based and x-vector-based MTEAD under both \textbf{Threshold} and \textbf{Oracle} conditions, and larger improvements are observed under the \textbf{Oracle} condition. We speculate that this is because when the number of speakers is correct in the initial diarization, the model can estimate a more accurate z-vector for each speaker. In contrast, under the \textbf{Threshold} condition, the incorrect number of speakers predicted in x-vector/AHC may cause some z-vectors to not match actual speakers, resulting in less reductions in DER and JER. The results in Tables \ref{tab:CallHome_Result_TS} and \ref{tab:Callhome_Result_all} show that, with its well-designed Extractor and Detector, MTEAD is a flexible and effective SD model that can extract more accurate speaker representations and handle conversations with different numbers of speakers. Compared to x-vector/AHC, x-vector/AHC+VB, and SA-EEND-EDA, MTEAD with z-vector achieved relative DER reductions of 30.9\%, 26.7\%, and 6.4\%, respectively.

Finally, we compared MTEAD with other models. Since most NN-based SD models were only evaluated in 2-speaker experiments, we compared different models on the 2-speaker CALLHOME task. From Table \ref{tab:CallHome_2speaker}, MTEAD outperformed all models except DIVE \cite{Zeghidour2021}, with a 12.2\% relative reduction in DER compared to x-vector/AHC \cite{Horiguchi2020}. According to Tables \ref{tab:Callhome_Result_all} and \ref{tab:CallHome_2speaker}, MTEAD outperformed most of the models compared in this study in both 2-speaker and multi-speaker tasks. This study focuses on improving the shortcomings of TS-VAD. Although there are many other advanced end-to-end or unsupervised SD models \cite{Dissen2022, Kinoshita2022}, we did not include them in the comparison due to the different training conditions of these models and paper length constraints. 

\vspace{-5pt}
\section{Conclusion}
\vspace{-5pt}
We have proposed the MTEAD model for speaker diarization, which consists of a Time-speaker Contextualizer based detector and a z-vector extractor. Its detector allows MTEAD to handle conversations with varying numbers of speakers and to use data with any number of speakers during training. This detector addresses the weaknesses of TS-VAD while retaining its strengths. The dedicated z-vector extractor for speaker diarization also improves performance compared to traditional i-vector and x-vector methods. The experimental datasets used in this study are relatively clean. We will try more challenging tasks such as DIHARD or CHiME in the future.
\newpage
\bibliographystyle{IEEEtran}
\bibliography{references.bib}

\begin{thebibliography}{10}
\providecommand{\url}[1]{#1}
\csname url@samestyle\endcsname
\providecommand{\newblock}{\relax}
\providecommand{\bibinfo}[2]{#2}
\providecommand{\BIBentrySTDinterwordspacing}{\spaceskip=0pt\relax}
\providecommand{\BIBentryALTinterwordstretchfactor}{4}
\providecommand{\BIBentryALTinterwordspacing}{\spaceskip=\fontdimen2\font plus
\BIBentryALTinterwordstretchfactor\fontdimen3\font minus
  \fontdimen4\font\relax}
\providecommand{\BIBforeignlanguage}[2]{{%
\expandafter\ifx\csname l@#1\endcsname\relax
\typeout{** WARNING: IEEEtran.bst: No hyphenation pattern has been}%
\typeout{** loaded for the language `#1'. Using the pattern for}%
\typeout{** the default language instead.}%
\else
\language=\csname l@#1\endcsname
\fi
#2}}
\providecommand{\BIBdecl}{\relax}
\BIBdecl

\bibitem{Anguera2010}
X.~Anguera, S.~Bozonnet, N.~Evans, C.~Fredouille, and G.~Friedland, ``{Speaker
  diarization: A review of recent research},'' \emph{IEEE Transactions on
  Audio, Speech and Language Processing}, vol.~20, no.~2, pp. 356--370, 2012.

\bibitem{Ryant2021}
N.~Ryant, P.~Singh, V.~Krishnamohan, R.~Varma, K.~Church, C.~Cieri, J.~Du,
  S.~Ganapathy, and M.~Liberman, ``{The third DIHARD diarization challenge},''
  in \emph{Proc. Interspeech}, 2021.

\bibitem{Diez2018a}
M.~Diez, L.~Burget, and P.~Matejka, ``{Speaker diarization based on bayesian
  HMM with eigenvoice priors},'' in \emph{Proc. Odyssey}, 2018.

\bibitem{Fujita2019}
Y.~Fujita, N.~Kanda, S.~Horiguchi, K.~Nagamatsu, and S.~Watanabe, ``{End-to-end
  neural speaker diarization with permutation-free objectives},'' in
  \emph{Proc. Interspeech}, 2019.

\bibitem{Snyder2018}
D.~Snyder, D.~Garcia-Romero, G.~Sell, D.~Povey, and S.~Khudanpur, ``{X-vectors:
  Robust DNN embeddings for speaker recognition},'' in \emph{Proc. ICASSP},
  2018.

\bibitem{Daniel2017}
D.~Garcia-Romero, D.~Snyder, G.~Sell, D.~Povey, and A.~McCree, ``{Speaker
  diarization using deep neural network embeddings},'' in \emph{Proc. ICASSP},
  2017.

\bibitem{Diez2019}
M.~Diez, L.~Burget, S.~Wang, J.~Rohdin, and H.~{\v{C}}ernock{\'{y}},
  ``{Bayesian HMM based x-vector clustering for speaker diarization},'' in
  \emph{Proc. Interspeech}, 2019.

\bibitem{Sell2018}
G.~Sell, D.~Snyder, A.~Mccree, D.~Garcia-Romero, J.~Villalba, M.~Maciejewski,
  V.~Manohar, N.~Dehak, D.~Povey, S.~Watanabe, and S.~Khudanpur, ``{Diarization
  is hard: Some experiences and lessons learned for the JHU team in the
  inaugural DIHARD challenge},'' in \emph{Proc. Interspeech}, 2018.

\bibitem{Dehak2011}
N.~Dehak, P.~J. Kenny, R.~Dehak, P.~Dumouchel, and P.~Ouellet, ``{Front-end
  factor analysis for speaker verification},'' \emph{IEEE Transactions on
  Audio, Speech and Language Processing}, vol.~19, pp. 788--798, 2011.

\bibitem{Sell2014}
G.~Sell and D.~Garcia-Romero, ``{Speaker diarization with PLDA i-vector scoring
  and unsupervised calibration},'' in \emph{Proc. IEEE SLT}, 2014.

\bibitem{Wan2018}
L.~Wan, Q.~Wang, A.~Papir, and I.~L. Moreno, ``{Generalized end-to-end loss for
  speaker verification},'' in \emph{Proc. ICASSP}, 2018.

\bibitem{Zhang2018}
A.~Zhang, Q.~Wang, Z.~Zhu, J.~Paisley, and C.~Wang, ``{Fully supervised speaker
  diarization},'' in \emph{Proc. ICASSP}, 2018.

\bibitem{Ioffe2006}
S.~Ioffe, ``{Probabilistic linear discriminant analysis},'' in \emph{Proc.
  ECCV}, 2006.

\bibitem{Simon2007}
S.~J. Prince and J.~H. Elder, ``{Probabilistic linear discriminant analysis for
  inferences about identity},'' in \emph{Proc. IEEE ICCV}, 2007.

\bibitem{Kenny2013}
P.~Kenny, T.~Stafylakis, P.~Ouellet, M.~Jahangir~Alam, and P.~Dumouchel,
  ``{PLDA for speaker verification with utterances of arbitary duration},'' in
  \emph{Proc. ICASSP}, 2013.

\bibitem{Ning2006}
H.~Ning, M.~Liu, H.~Tang, and T.~Huang, ``{A spectral clustering approach to
  speaker diarization},'' in \emph{Proc. Interspeech}, 2006.

\bibitem{Park2020}
T.~J. Park, K.~J. Han, J.~Huang, X.~He, B.~Zhou, P.~Georgiou, and S.~Narayanan,
  ``{Speaker diarization with lexical information},'' in \emph{Proc.
  Interspeech}, 2019.

\bibitem{Bozonnet2010}
S.~Bozonnet, N.~W. Evans, and C.~Fredouille, ``{The lia-eurecom RT'09 speaker
  diarization system: Enhancements in speaker modelling and cluster
  purification},'' in \emph{Proc. ICASSP}, 2010.

\bibitem{Shum2013}
S.~H. Shum, N.~Dehak, R.~Dehak, and J.~R. Glass, ``{Unsupervised methods for
  speaker diarization: An integrated and iterative approach},'' \emph{IEEE
  Transactions on Audio, Speech and Language Processing}, vol.~21, no.~10, pp.
  2015--2028, 2013.

\bibitem{Sell2015}
G.~Sell and D.~Garcia-Romero, ``{Diarization resegmentation in the factor
  analysis subspace},'' in \emph{Proc. ICASSP}, 2015.

\bibitem{Sahidullah2019}
\BIBentryALTinterwordspacing
M.~Sahidullah, J.~Patino, S.~Cornell, R.~Yin, S.~Sivasankaran, H.~Bredin,
  P.~Korshunov, A.~Brutti, R.~Serizel, E.~Vincent, N.~Evans, S.~Marcel,
  S.~Squartini, and C.~Barras, ``{The speed submission to DIHARD II:
  Contributions {\&} lessons learned},'' 2019. [Online]. Available:
  \url{http://arxiv.org/abs/1911.02388}
\BIBentrySTDinterwordspacing

\bibitem{Fujita2019b}
\BIBentryALTinterwordspacing
Y.~Fujita, S.~Watanabe, S.~Horiguchi, Y.~Xue, and K.~Nagamatsu, ``{End-to-End
  neural diarization: Reformulating speaker diarization as simple multi-label
  classification},'' 2020. [Online]. Available:
  \url{http://arxiv.org/abs/2003.02966}
\BIBentrySTDinterwordspacing

\bibitem{Neumann2018}
T.~von Neumann, K.~Kinoshita, M.~Delcroix, S.~Araki, T.~Nakatani, and
  R.~Haeb-Umbach, ``{All-neural online source separation, counting, and
  diarization for meeting analysis},'' in \emph{Proc. ICASSP}, 2018.

\bibitem{Fujita2019d}
Y.~Fujita, N.~Kanda, S.~Horiguchi, K.~Nagamatsu, and S.~Watanabe, ``{End-to-end
  neural speaker diarization with permutation-free objectives},'' in
  \emph{Proc. Interspeech}, 2019.

\bibitem{Bullock2020}
L.~Bullock, H.~Bredin, and L.~P. Garcia-Perera, ``{Overlap-aware diarization:
  Resegmentation using neural end-to-end overlapped speech detection},'' in
  \emph{Proc. ICASSP}, 2020.

\bibitem{Yu2017}
D.~Yu, M.~Kolb{\ae}k, Z.-H. Tan, and J.~Jensen, ``{Permutation invariant
  training of deep models for speaker-independent multi-talker speech
  separation},'' in \emph{Proc. ICASSP}, 2017.

\bibitem{Zeghidour2021}
N.~Zeghidour, O.~Teboul, and D.~Grangier, ``{DIVE: End-to-end speech
  diarization via iterative speaker embedding},'' in \emph{Proc. IEEE ASRU},
  2021.

\bibitem{Fujita2019c}
Y.~Fujita, N.~Kanda, S.~Horiguchi, Y.~Xue, K.~Nagamatsu, and S.~Watanabe,
  ``{End-to-end neural speaker diarization with self-attention},'' in
  \emph{Proc. IEEE ASRU}, 2019.

\bibitem{Horiguchi2020}
S.~Horiguchi, Y.~Fujita, S.~Watanabe, Y.~Xue, and K.~Nagamatsu, ``{End-to-end
  speaker diarization for an unknown number of speakers with encoder-decoder
  based attractors},'' in \emph{Proc. Interspeech}, 2020.

\bibitem{Horiguchi2021}
S.~Horiguchi, P.~Garcia, Y.~Fujita, S.~Watanabe, and K.~Nagamatsu,
  ``{End-to-end speaker diarization as post-processing},'' in \emph{Proc.
  ICASSP}, 2021.

\bibitem{Medennikov2020a}
I.~Medennikov, M.~Korenevsky, T.~Prisyach, Y.~Khokhlov, M.~Korenevskaya,
  I.~Sorokin, T.~Timofeeva, A.~Mitrofanov, A.~Andrusenko, I.~Podluzhny,
  A.~Laptev, and A.~Romanenko, ``{Target-speaker voice activity detection: A
  novel approach for multi-speaker diarization in a dinner party scenario},''
  in \emph{Proc. Interspeech}, 2020.

\bibitem{Luo2020}
Y.~Luo, Z.~Chen, and T.~Yoshioka, ``{Dual-path RNN: Efficient long sequence
  modeling for time-domain single-channel speech separation},'' in \emph{Proc.
  ICASSP}, 2020.

\bibitem{Chenda2021}
C.~Li, Y.~Luo, C.~Han, J.~Li, T.~Yoshioka, T.~Zhou, M.~Delcroix, K.~Kinoshita,
  C.~Boeddeker, Y.~Qian, S.~Watanabe, and Z.~Chen, ``{Dual-path RNN for long
  recording speech separation},'' in \emph{Proc. IEEE SLT}, 2021.

\bibitem{Wang2022}
W.~Wang, Q.~Lin, D.~Cai, and M.~Li, ``{Similarity measurement of segment-level
  speaker embeddings in speaker diarization},'' \emph{IEEE/ACM Transactions on
  Audio Speech and Language Processing}, vol.~30, pp. 2645--2658, 2022.

\bibitem{Okabe2018}
K.~Okabe, T.~Koshinaka, and K.~Shinoda, ``{Attentive statistics pooling for
  deep speaker embedding},'' \emph{Proc. Interspeech}, 2018.

\bibitem{Hamers1989}
L.~Hamers, Y.~Hemeryck, G.~Herweyers, M.~Janssen, H.~Keters, R.~Rousseau, and
  A.~Vanhoutte, ``{Similarity measures in scientometric research: The Jaccard
  index versus Salton's cosine formula},'' \emph{Information Processing and
  Management}, vol.~25, no.~3, pp. 315--318, 1989.

\bibitem{Real1996}
R.~Real and J.~M. Vargas, ``{The probabilistic basis of Jaccard's index of
  similarity},'' \emph{Systematic Biology}, vol.~45, no.~3, pp. 380--385, 1996.

\bibitem{Ryant2019}
N.~Ryant, K.~Church, C.~Cieri, A.~Cristia, J.~Du, S.~Ganapathy, and
  M.~Liberman, ``{The second DIHARD diarization challenge: Dataset, task, and
  baselines},'' in \emph{Proc. Interspeech}, 2019.

\bibitem{Vaswani2017}
A.~Vaswani, N.~Shazeer, N.~Parmar, J.~Uszkoreit, L.~Jones, A.~N. Gomez,
  L.~Kaiser, and I.~Polosukhin, ``{Attention is all you need},'' in \emph{Proc.
  NIPS}, 2017.

\bibitem{Huang2020}
Z.~Huang, S.~Watanabe, Y.~Fujita, P.~Garc{\'{i}}a, Y.~Shao, D.~Povey, and
  S.~Khudanpur, ``{Speaker diarization with region proposal network},'' in
  \emph{Proc. ICASSP}, 2020.

\bibitem{Dissen2022}
Y.~Dissen, F.~Kreuk, and J.~Keshet, ``{Self-supervised speaker diarization},''
  in \emph{Proc. Interspeech}, 2022.

\bibitem{Kinoshita2022}
K.~Kinoshita, M.~Delcroix, and T.~Iwata, ``{Tight integration of neural- and
  clustering-based diarization through deep unfolding of infinite Gaussian
  mixture model},'' in \emph{Proc. ICASSP}, 2022.

\end{thebibliography}

\end{document}